\documentclass{PoS/PoS}
\usepackage{amsmath,url,cite}
\newcommand\MSB{\ensuremath{\overline{\rm MS}}}
\newcommand\mMSB{\ensuremath{m_{\scriptscriptstyle \overline{\rm MS}}}}
\newcommand\mpole{\ensuremath{m_{\scriptscriptstyle P}}}
\newcommand\mathd{{\rm d}}
\newcommand\pt{\ensuremath{p_{\rm \scriptscriptstyle T}}}
\newcommand\as{\ensuremath{\alpha_{\scriptscriptstyle \rm S}}}
\newcommand\lambdaQCD{\ensuremath{\Lambda_{\scriptscriptstyle \rm QCD}}}
\newcommand\LambdaQCD{\lambdaQCD}
\newcommand\POWHEG{{\tt POWHEG}}
\newcommand\HERWIG{{\tt HERWIG}}

\title{Theory Summary}

\ShortTitle{Theory Summary}

\author{\speaker{Paolo Nason}\\
        INFN, sez. di Milano Bicocca\\
        E-mail: \email{paolo.nason@mib.infn.it}}

\abstract{I review few selected topics on recent theoretical progress
in top physics. In particular I will discuss recent progress in the computation
of the relationship between the \MSB{} and pole top mass, in the NNLO calculation of top
differential distributions, and in the simulation of top
production and decays. Implications for top mass measurements will be discussed.}

\FullConference{8th International Workshop on Top Quark Physics, TOP2015\\
		14-18 September, 2015\\
		Ischia, Italy}

\begin{document}
\section{Introduction}
In the current year there has been considerable progress in
theoretical studies regarding top quark physics.  In the present
workshop, summaries of recent developments in several areas have been
presented.  As reminded in the introductory talk~\cite{Peskin}, in
several extensions of the Standard Model, the large mass of the top
quark plays an important role in driving the Electro-Weak symmetry
breaking. It is thus natural to expect new physics effects to show up
in processes involving the top. It was also reminded~\cite{Espinosa}
that, having the largest Yukawa coupling, the
top can destabilize the Higgs potential.
It is thus important to explore top physics in great detail,
both with regard to searches for New Physics signals, and with regard
to precision measurements of its properties, in particular of its
mass.

At the LHC, top production is a very abundant process, thus offering
new opportunities for the search of new physics effects in flavour
violating top decays~\cite{Schulze} and in anomalous
couplings~\cite{Schulze,DeAndrea,Sharma}.
Signals of new physics are expected to
become stronger for highly boosted tops. The corresponding reduction
in rates can be compensated by the use of boosted top techniques, that
allow for tagging hadronic top decays~\cite{Spannowsky}.  In
general, all these studies require a quite precise understanding of
the production and decay properties of the top in the Standard model.
Several recent results have appeared on the computation of higher order
QCD corrections to top production and decays, including their matching
to parton shower generators~\cite{Heymes,Papanastasiou,Pecjak,Re}, and of EW
corrections to various top processes~\cite{Zaro,Uwer,Pagani}.
Theoretical issues in the measurement of the top mass have been
summarized in~\cite{Corcella}.

It is impossible to summarize all the material presented in this
workshop, especially in view of the fact that several talks were
already summaries on their own.  Here I will focus in detail upon
three topics where very recent and important progress was made: the
relationship between the \MSB{} and pole top mass, and its
implications for top mass measurements; the first
presentation of NNLO distributions for $t\bar{t}$ production, that was
given by D.~Heymes at this conference~\cite{Heymes}; and finally
I will discuss the first implementations of
NLO calculation matched with shower generators (NLO+PS) that can
handle properly finite width effects in the cases when resonances can
decay into coloured particles, and its application to top production
processes.

In the conclusions, I will discuss in particular the relevance of this
recent progress for the top mass measurement at the LHC.

\section{The \MSB{} and pole mass relation for the top}
The top mass term in the QCD lagrangian arises from the Yukawa
coupling of the top to the Higgs field.
Mass renormalization can be naturally carried out with the
same methods adopted for all other couplings, i.e. the \MSB{}
scheme. Alternatively we can define the mass parameter as the particle physical mass, i.e. its
rest energy. This parameter is also called the ``pole mass'', since it
really corresponds to the pole position in the particle propagator.
This definition is sometimes more convenient, in view of the fact that
the physical mass is often a parameter that can be directly measured, and that
it remains fixed as we perform higher order calculations. If we insist
in using the \MSB{} mass, on the other hand, we would find that, as we
raise the perturbative order, the expression of the physical mass
receives higher order corrections.

In case of stable coloured massive
particles, there are practical and theoretical obstacles in
defining the pole mass. Such particles, because of colour confinement, cannot
be isolated. We thus expect to find them in bound states.
Intuitively, we expect that the mass of these bound states will differ from the
pole mass by corrections
of order $\LambdaQCD$.  It is remarkable that this argument,
relying upon color confinement, is also supported by
perturbation theory alone, that suggests that a similar ambiguity in the pole
mass should also arise due to self-energy
corrections. The argument can be summarized as follows. It can be shown that
infrared gluons in a heavy quark self-energy
yield mass corrections
of the form
\begin{equation}\label{eq:selfen}
\delta m \propto \int_0^m \mathd l\; \as,
\end{equation}
where $l$ is the gluon virtuality. This expression is infrared finite, but one should not
forget that the inclusion of higher order corrections leads to the running of the strong coupling constant,
i.e. to the replacement,
\begin{equation}\label{eq:alpharun}
\as \to \as(l^2) = \frac{1}{b_0\log l^2/\lambdaQCD^2} = \frac{\as(m^2)}{1-b_0\as(m^2) \log m^2/l^2},
\end{equation}
that leads to a divergence in the integral when $l^2=\lambdaQCD^2$.

If we expand eq.~\ref{eq:alpharun}
in powers of $\as(m^2)$ we get
\begin{equation}\label{eq:alpharunexp}
\as \to \as(l^2) = \as(m^2) \sum_{k=1}^\infty \left(b_0\as(m^2) \log m^2/l^2\right)^k.
\end{equation}
Replacing this expansion in eq.~\ref{eq:selfen} we get
\begin{equation}\label{eq:selfenexp}
\delta m \propto \int^m \mathd l\; \as(l^2)=\;\as(m^2) \sum_{k=1}^\infty \left(2b_0\as(m^2)\right)^k \int^m \mathd l
\; \log^k \frac{m}{l} \propto m\;\sum_{k=1}^\infty \left(2 b_0\as(m^2)\right)^k (k-1)!,
\end{equation}
so that, even if each term of the expansion is finite, the sum is
divergent because of the factorially growing coefficients. This sums
are
known as infrared renormalons (see ref.~\cite{Beneke:1998ui} for a
review). The name reminds us of the infrared nature of the problem
(i.e. the small values of momenta involved) and from its relation to
the running coupling.

From eq.~(\ref{eq:alpharun}) we are led to conclude that there is an
intrinsic ambiguity of order \LambdaQCD{} in the heavy quark pole
mass. In fact, when $l \approx \lambdaQCD$, the coupling constant
becomes of order one, and we are no longer allowed to use the
perturbative expansion, so that we are unable to compute the integral
in the region $l \lesssim \lambdaQCD$.  A similar conclusion follows
if we use the expansion of eq.~(\ref{eq:alpharunexp}) as an asymptotic
expansion, stopping the summation when the terms of the series stop decreasing.
This happen when
\begin{equation}
\frac{(2 b_0)^{n+1} \as^{n+1} n!}{(2 b_0)^n \as^n (n-1)!}\approx 2 b_0 \as n \Rightarrow 1\,
\end{equation}
that is to say when $n \approx 1/(2 b_0 \as)$. The size of the last term at this value
of $n$ is
\begin{equation}
(2 b_0)^{n+1} \as^{n+1} n!\approx n^{-n} (n^n e^{-n}) \approx
  \exp\left(-\frac{1}{2b_0 \as}\right) \approx \frac{\LambdaQCD}{m},
\end{equation}
yielding again a mass ambiguity of order $\LambdaQCD$.

The presence of infrared renormalons in the relation between the \MSB{} and pole mass of
heavy quarks was first pointed out it ref.~\cite{Beneke:1994sw}. In particular, in ref.~\cite{Beneke:1994rs}
it was found that the structure of the renormalon has the form
\begin{eqnarray} \label{eq:beneke2}
\mpole &=& \mMSB \left(1+\sum_{n=0}^\infty r_n \alpha_s^{n+1} \right)
r_n \to N (2 b_0)^n\, \Gamma(n+1+b) \left(1+\sum_{k=1}^\infty \frac{s_k}{n^k}\right),
\end{eqnarray}
where $b_0$ and $b_1$ are coefficients of the $\beta$ function according to the definition
\begin{equation}
\mu^2 \frac{\partial \alpha_s}{\partial \mu^2} = -b_0 \as^2-b_1\as^3 \,\ldots
\end{equation}
and the $s_i$ coefficients can all be computed in terms of the coefficients of the
$\beta$ function. The normalization $N$ cannot be computed with present techniques.

It is often stated that, if the top pole mass is measured, an intrinsic error of order
of few hundred MeV is to be expected, and should thus be ascribed to all LHC mass
measurement that rely on observables strongly sensitive to the mass of the system comprising
the top the decay products.

\begin{figure}
\begin{center}
\includegraphics[width=8cm]{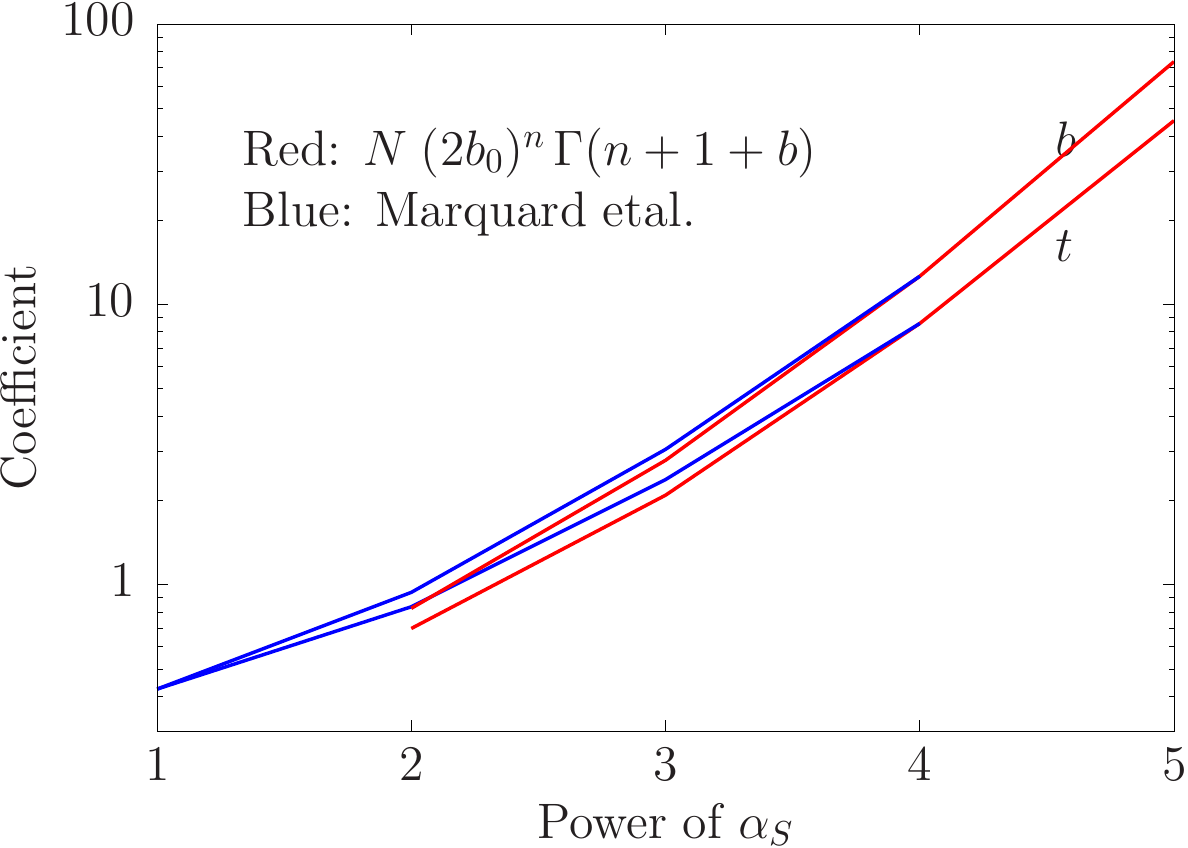}
\caption{Fit of the coefficient of the renormalon approximation to the exact one,
for both the bottom and the top quark. The normalization $N$ of the renormalon
formula is fixed to match the exact 4th order coefficient.}
\label{fig:mpolemsbartb}
\end{center}
\end{figure}

In the current year, in ref.~\cite{Marquard:2015qpa} (from now on MSSS)
remarkable progress was made in the calculation of the relation
between the \MSB{} and pole mass of a heavy quark, reaching now 4-loop accuracy.
For the top, this reads
\begin{equation}\label{eq:marquard1}
\mpole = \mMSB (1 + 0.4244\as + 0.8345\as^2 + 2.375\as^3 + (8.49 \pm 0.25)\as^4)
\end{equation}
where \mMSB{} is the top \MSB{} mass evaluated at a scale equals to itself.
The strong coupling \as{} is the 6-flavour strong coupling, also evaluated at \mMSB.
The corresponding corrections to the top mass (assuming $\mMSB=163.643\;$GeV) are given by
\begin{equation}\label{eq:marquard2}
\mpole = 163.643+7.557+1.617+0.501 + 0.195\,{\rm GeV},
\end{equation}
showing that the perturbative expansion is still well behaved, the last term being
less than $1/2$ of the previous term. We could just take this as an indication
that the intrinsic error on the pole mass is less than 200~MeV. It is possible,
however, to do better than this if we recognize that the behaviour of the
perturbative expansion up to the fourth order is well-fitted by formula (\ref{eq:beneke2}).
This is shown in fig.~\ref{fig:mpolemsbartb}.
As we can see, by fitting the normalization $N$ in eq.~\ref{eq:beneke2} using the exact
value of the 4th order coefficient from the MSSS calculation,
the 3rd and 2nd order coefficients turn out to fit
the exact perturbative calculation quite well.\footnote{This fact was already studied
in $b$ physics contexts in~\cite{Pineda:2001zq,Ayala:2014yxa}.} Assuming thus that
the perturbative expansion is dominated by the renormalon for higher order terms,
we can guess the rest of the expansion. For example, the term
of order 5 in eq.~\ref{eq:marquard1} is $45.4 \as^5$, corresponding roughly to a $110$~MeV correction
in eq.~\ref{eq:marquard2}.
The approximate perturbative series reaches its minimum for $n \approx 8$, and the size of the last
term corresponds to a 66~MeV mass correction.\footnote{A more detail analysis of this issue
is in preparation~\cite{SteihauserXXX}.}

Summarizing, there is a strong suggestion that the renormalon problem
in the pole mass is not very severe, at least for the precision that the
LHC is aiming to.  This is good news for the standard LHC top mass
determinations~\cite{ACastro}, that make use of observables that are
intimately related to the pole mass.

\section{NNLO differential distributions for
$t\bar{t}$ pair production}
\begin{figure}
\begin{center}
\includegraphics[width=8cm]{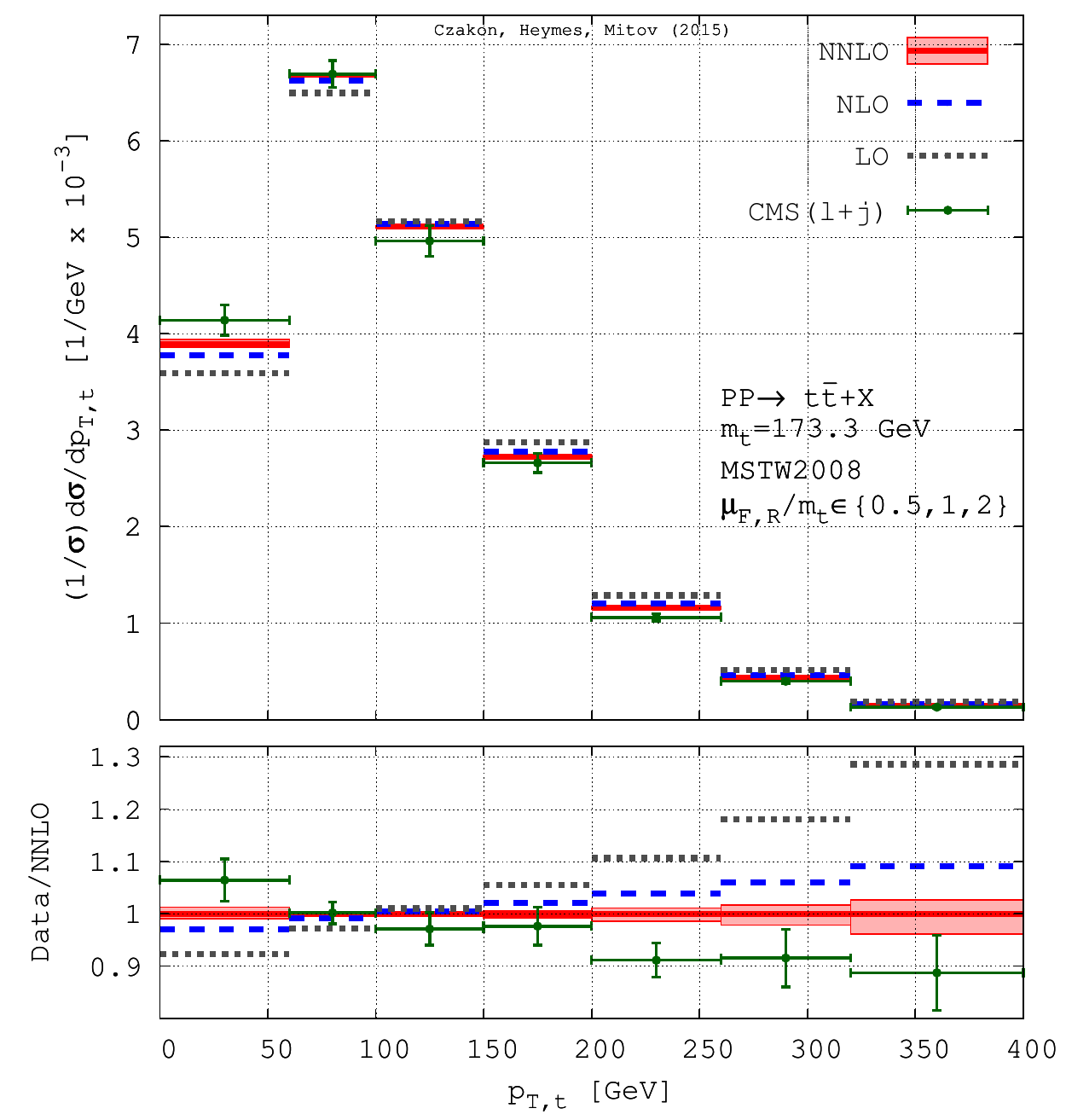}
\includegraphics[width=7cm]{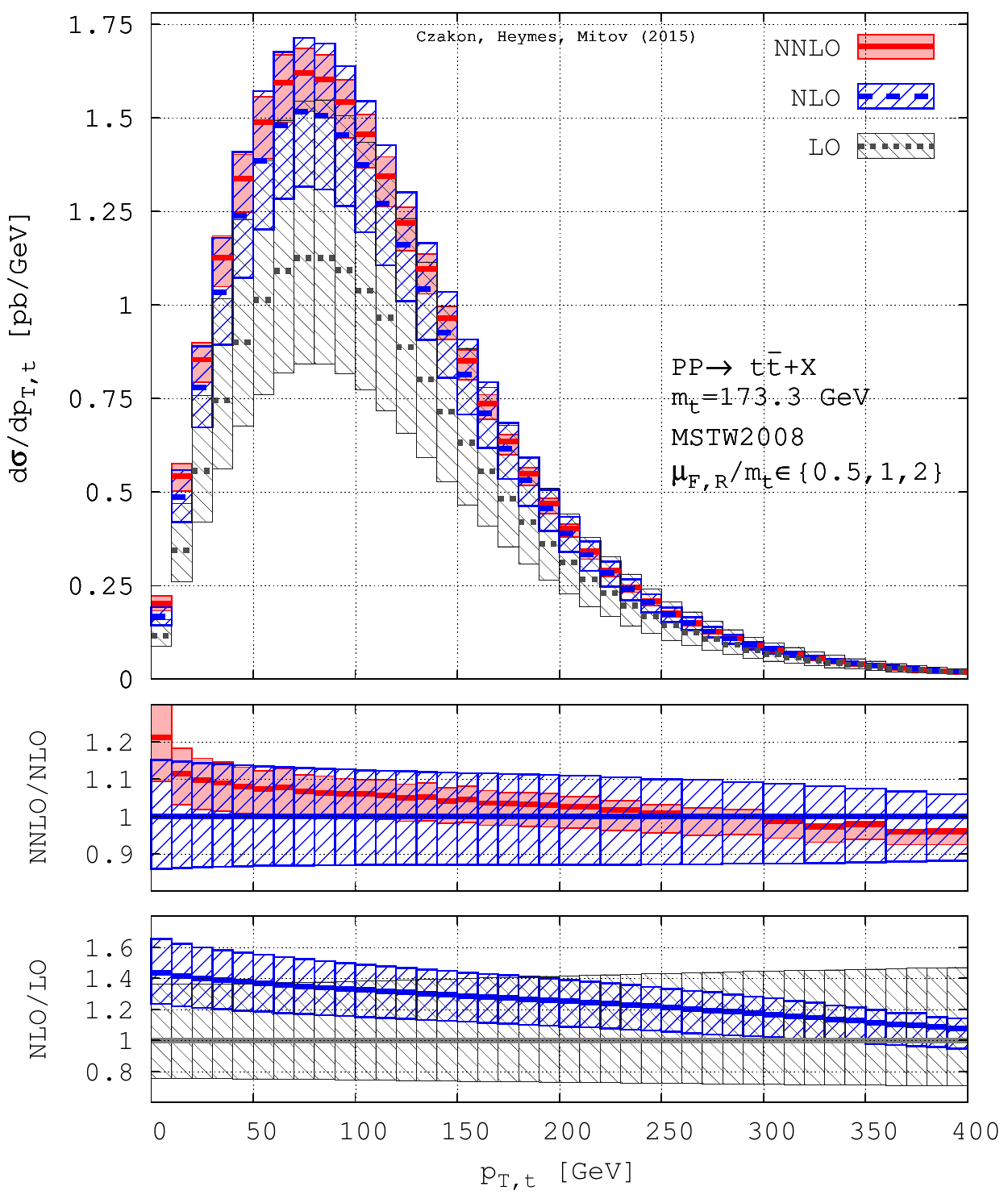}
\caption{NNLO results for the top transverse momentum distribution.}
\label{fig:tptCMS}
\end{center}
\end{figure}

D. Heymes has presented in this workshop the first NNLO differential
distributions for top pair production. Needless to say, this is
an outstanding results. A relative publication has
appeared~\cite{Czakon:2015owf}.
Figure~\ref{fig:tptCMS},
taken from~\cite{Czakon:2015owf}, reports NLO and NNLO results
for the top transverse momentum distribution, compared to CMS data.
As also shown by Hindrichs at this conference~\cite{OHindrichs},
this particular distribution is not very well described by available
LO+PS and NLO+PS generators, that also include
shower and finite width effects
(not present in the fixed order results of fig.~\ref{fig:tptCMS}).
Discrepancies show up at very low and very high top transverse momentum.
In particular, the measured cross section is roughly 10\%{} higher in the
first transverse momentum bin, and it is lower in the high
transverse momentum tail.
Keeping this mind, we may use fig.~\ref{fig:tptCMS} as a guidance
for what variation we may expect relative to an NLO calculations when
NNLO corrections are included.
The slope of the NNLO/NLO red band on the right plot of the figure seems to indicate
that the inclusion of higher order effects should alleviate
both the low and the high $\pt$  discrepancies.
It should be said, however, that the use of a fixed
renormalization and factorization scale is bound to overestimate
the cross section in the high transverse momentum region, where a scale
of the order of the top transverse mass (i.e. $\sqrt{\pt^2+m^2}$)
should be more appropriate, and is in fact used in LO+PS and NLO+PS
generators. In the last bin of the plot, the transverse
mass is roughly 400~GeV, not far from the high value of the scale
variation, that is twice the top mass. On the other hand, the right plot seems to
indicate that the NNLO result would not be lower than the NLO one if
the transverse mass was used as central scale, and that
the NNLO/NLO slope may disappear,
thus leading to no improvement in the comparison with data.\footnote{This point was
raised by M. L. Mangano at this conference.} Further numerical studies may
clarify this issue.

\begin{figure}[h]
\begin{center}
  \begin{minipage}{0.56\textwidth}
  \includegraphics[width=\textwidth]{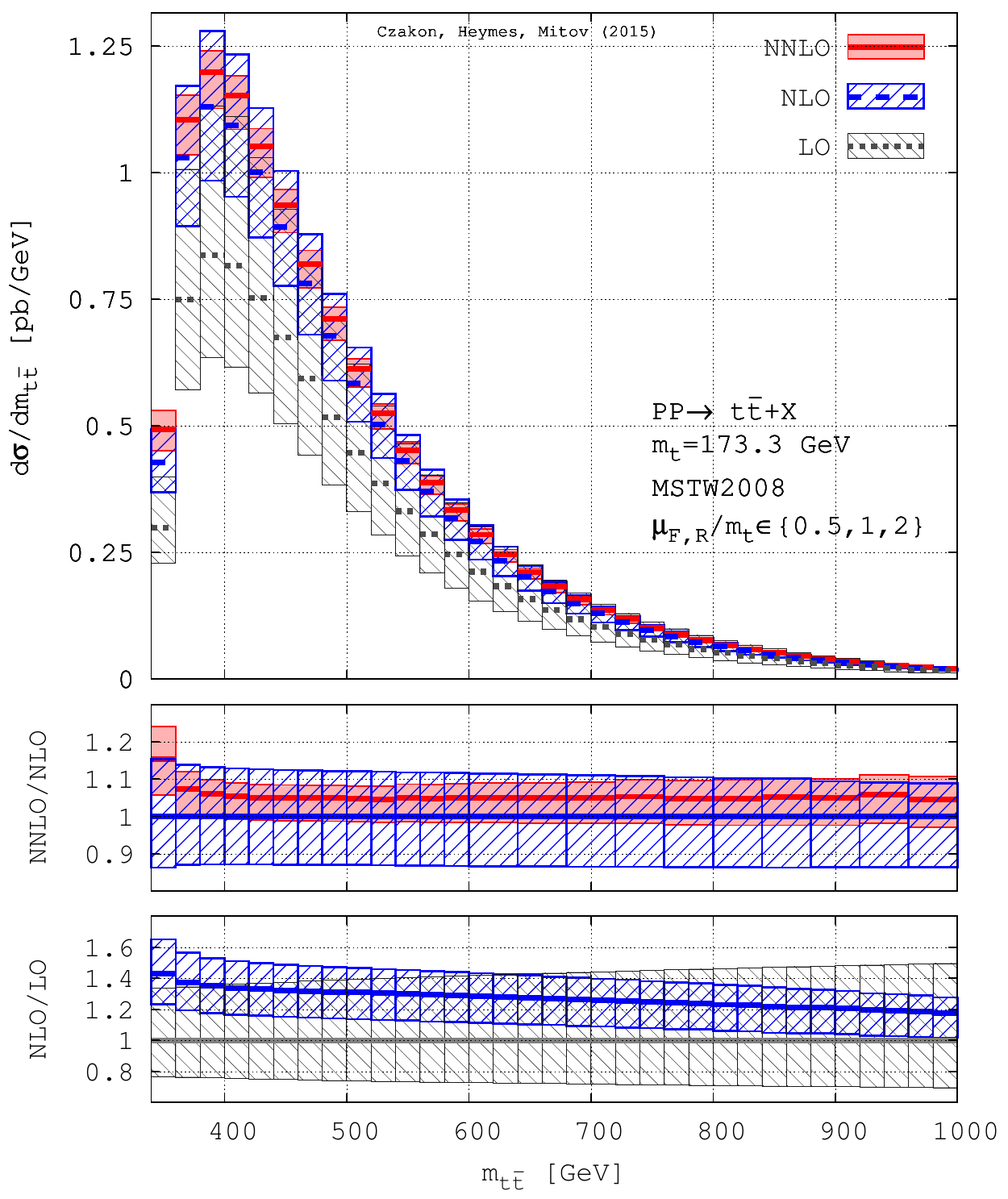}
  \end{minipage}
  \begin{minipage}{0.43\textwidth}
  \includegraphics[width=\textwidth]{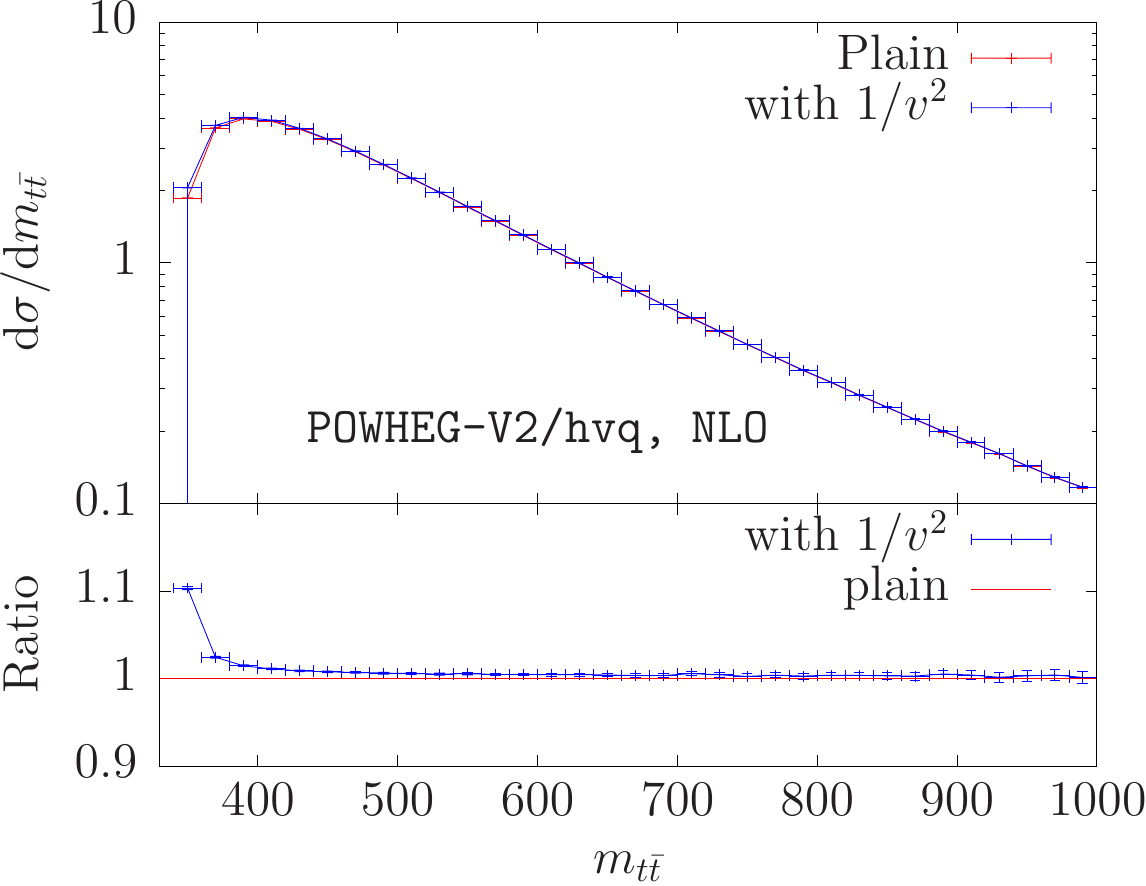}
  \includegraphics[width=\textwidth]{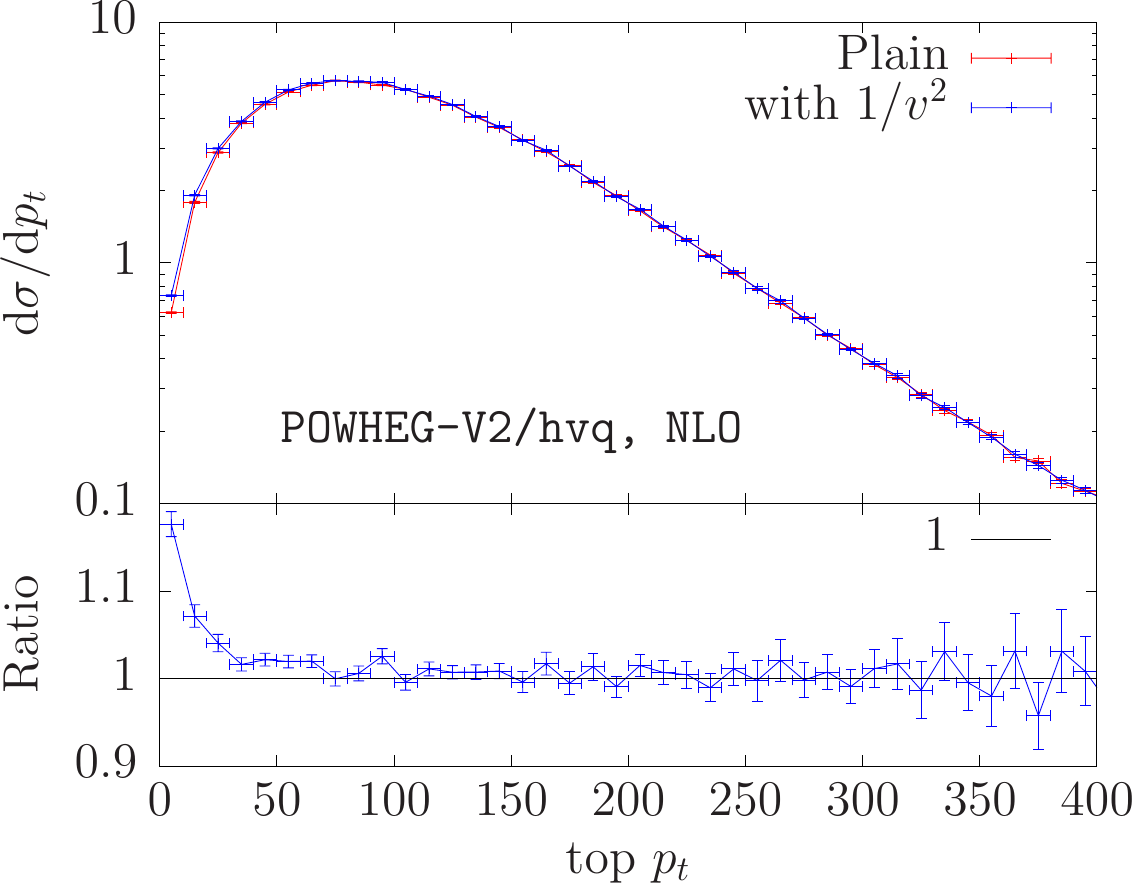}
  \end{minipage}
\caption{Left:  results for the invariant mass of the
  $t\bar{t}$ system. Right: Invariant mass of the $t\bar{t}$ pair and transverse momentum of the top obtained
    with the {\tt POWHEG-BOX-V2/hvq} generator, modified with the inclusion of the
    $1/v^2$ NLO term, compared to the plain result.}
\label{fig:ttmass}
\label{fig:pwhg-v2}
\end{center}
\end{figure}

The discrepancy in the low transverse momentum bin also deserves some
comment. It yields a useful example of the interplay of Monte Carlo
studies, higher order calculations and comparison with data.
First of all, we notice that also the
invariant mass of the $t\bar{t}$ pair, shown in
Fig.~\ref{fig:ttmass},
displays a discrepancy in the threshold region.
Within the framework of the CERN top working group, it was pointed out
in several circumstances that the NLO+PS generator that was yielding
the best description of the top transverse momentum spectrum was
\POWHEG{} interfaced with \HERWIG{}. As far as the threshold region
is concerned, this fact was eventually
traced back to the way in which momentum reshuffling is handled in
\HERWIG{}~\cite{TOPLHCWG-may-21-2014}. Since reshuffling in NLO+PS
generators yields corrections that are formally of NNLO order, it
was argued that until a fully differential NNLO result was available, the
reshuffling ambiguity should have been considered as a source of a
theoretical error.
An NNLO calculation is now available, and we can proceed to investigate
whether it justifies the \HERWIG{} reshuffling strategy.
From fig.~\ref{fig:tptCMS} it appears
that an enhancement at very low transverse momentum actually arises in
the NNLO result. However, before attributing it to
reshuffling effects, we should consider other sources of
enhancements that may appear at the NNLO level. New Sudakov logarithms
arise at the NNLO level, but they should also be reasonably modeled at the
leading logarithmic level by
shower Monte Carlo.
On the other hand, NNLO $1/v$ singularities (where $v$ is the top velocity
in the $t\bar{t}$ frame)
are not modeled at all in Monte Carlo generators. The coefficient of the
$1/v^2$ term at NNLO is quite large, having relative size
equal to $(\as/4\pi)^2 \times 68.5/v^2$ \cite{Beneke:2009ye}.
It is easy to implement such correction in the {\tt POWHEG-BOX-V2/hvq} package.
In fig.~\ref{fig:pwhg-v2} I display the effect of such modification for the
invariant mass of the $t\bar{t}$ pair and for the top transverse momentum.
As one can see, the effect in the low $p_t$ and low invariant mass region is quite compatible with
what is observed in the NNLO result. This hints to the fact that Coulomb effects, rather
than recoil, are responsible for the mismodeling of the low invariant mass region.
A detailed study of this problem is in preparation~\cite{Mitov}.

\section{NLO+PS simulation of top pair production and decay}
This section is about current progress in the simulation of heavy quark pair production.
NLO+PS generators for this process have been available for a long time~\cite{Frixione:2003ei,Frixione:2007nw},
and are today available also for associated production with jets~\cite{Re}.
Top decay is only treated approximately in these generators, with no inclusion of NLO corrections.
In view of the interest in an accurate determination of the top mass,
the inclusion of radiation in decay, and also of interference
effects, has been addressed in the literature.

In ref.~\cite{Campbell:2014kua}, an NLO-PS generator including
full spin correlation and radiation in decay was presented. This generator uses zero-width, factorized
matrix elements for heavy flavour production and decay. Interference effects in radiation are
thus not included. Corrections are however applied in order to account for finite width, non-resonant and interference
effects at least at the leading order. This generator has been illustrated in the talk by E. Re in these
proceedings~\cite{Re}.

\begin{figure}
\begin{center}
  \includegraphics[width=0.33\textwidth]{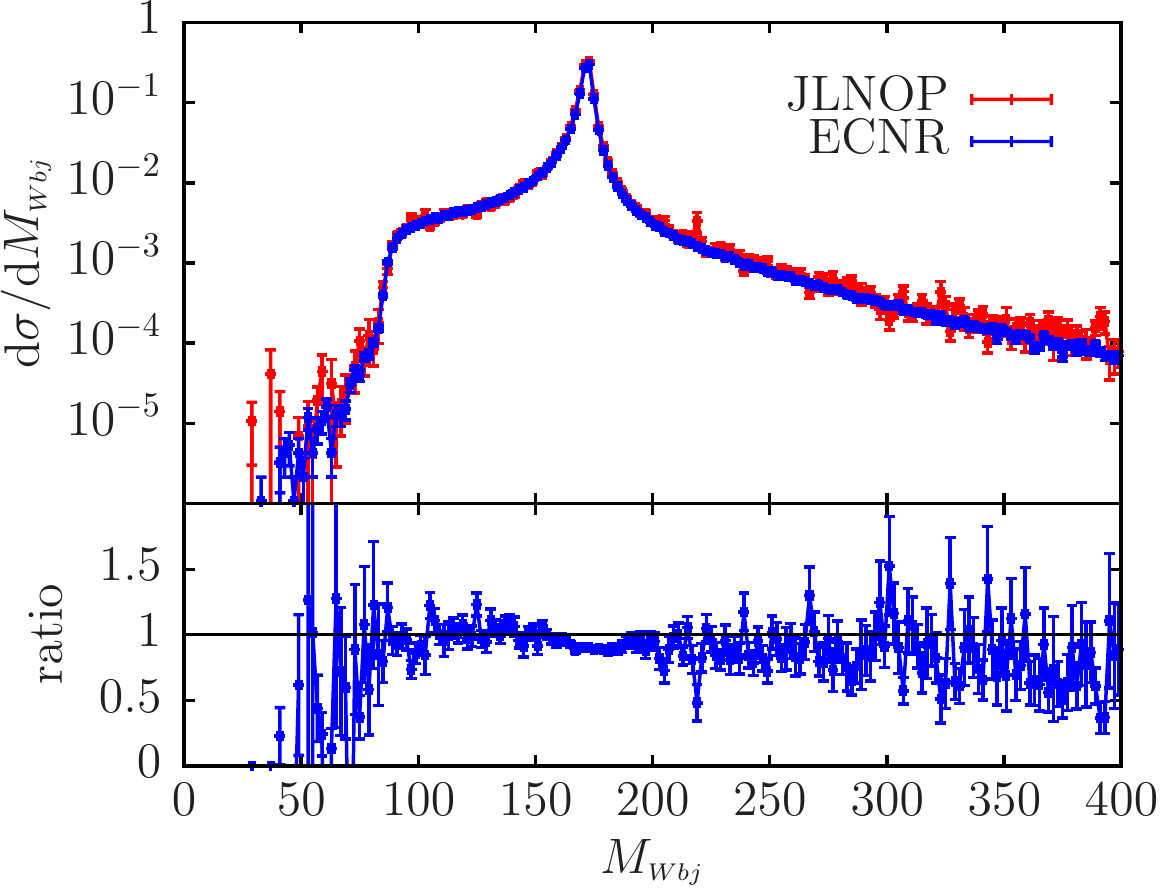}
  \includegraphics[width=0.32\textwidth]{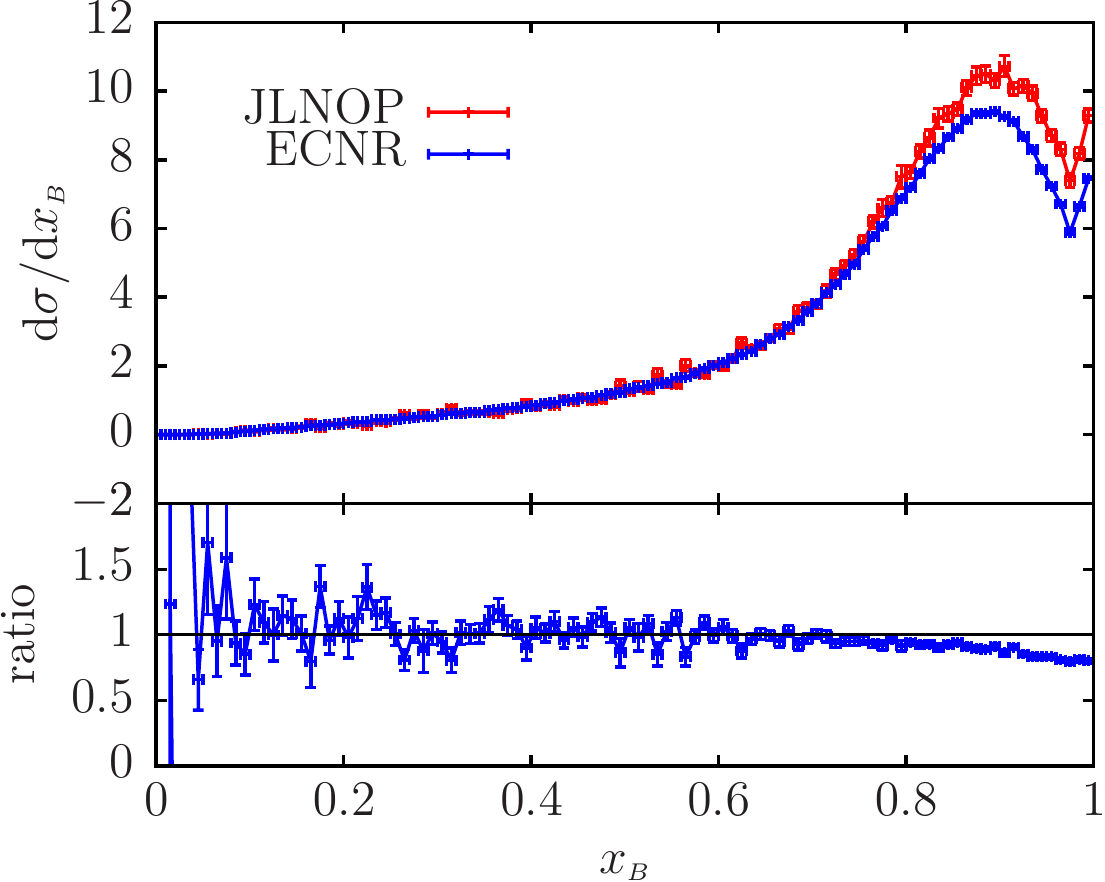}
  \includegraphics[width=0.33\textwidth]{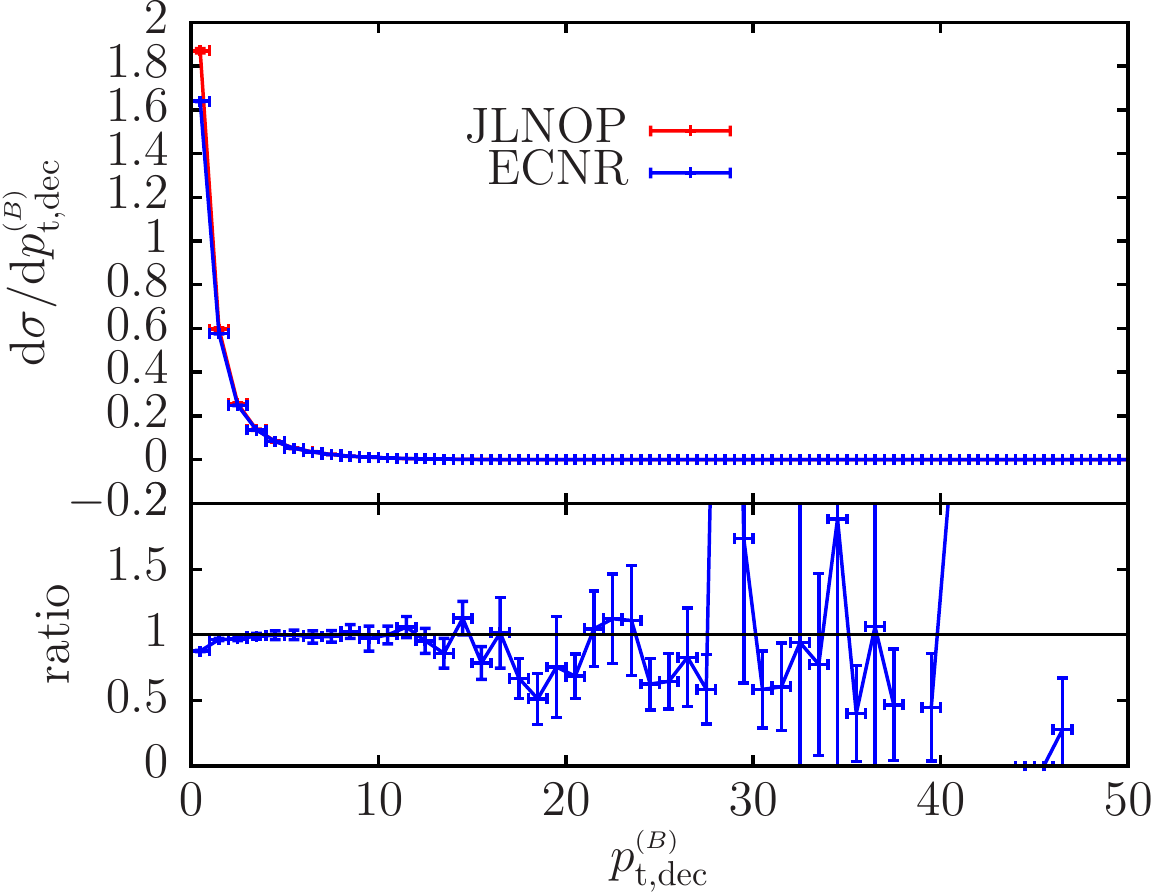}
  \caption{
    Comparison of the ECNR (for Ellis-Campbell-Nason-Re) generator with respect to the JLNOP (for Jezo-Lindert-Oleari-Nason-Pozzorini)
    one for $W^{+} W^{-} b \bar{b}$ production. The left plot represents the invariant mass distribution of the $W^+$-$\; b_{\rm jet}$ system,
    the central plot represents the fragmentation function of the $B$ meson defined in the $W^+$-$\;b_{\rm jet}$ centre-of-mass,
    and the right plot represents the transverse momentum of the $B$ meson with respect to the recoiling $W^+$ boson
    in the $W^+$-$\;b_{\rm jet}$ centre-of-mass.}
  \label{fig:ttbprodanddec}
\end{center}
\end{figure}

In ref.~\cite{Garzelli:2014dka}, an NLO+PS generator for the full process $W^{+} W^{-} b \bar{b}$ was
built in the
so called {\tt POWHEL} framework (consisting essentially of the {\tt POWHEG-BOX-V2} interfaced
to the {\tt HELAC} matrix elements).
The {\tt POWHEG-BOX-V2} framework does not treat resonances in any special way, i.e. it
generates radiation from the top or anti-top decay product
without accounting for the fact that near the resonance peak this radiation should preserve the mass of the
resonances. It is not clear at the moment whether this is causing any mismodeling in practice.
It is important, however, to develop a correct treatment of resonances in the NLO+PS framework,
since it can be proven that standard NLO+PS generators, in the narrow width limit,
will develop a  distorsion of the jet-mass spectrum when $m_{\rm jet}\approx \Gamma E_{\rm jet}$,
where $\Gamma$ is the resonance width, and $m_{\rm jet}$, $E_{\rm jet}$ are respectively the jet
mass and its energy~\cite{Jezo:2015aia}.
A formalism for such treatment has been put forward in ref.~\cite{Jezo:2015aia}, where it was applied to the
process of $t$-channel single top production. In this conference, an analogous effort in the
{\tt MadGraph5\_aMC@NLO} framework was presented~\cite{Papanastasiou}.
An NLO+PS generator for $W^{+} W^{-} b \bar{b}$ production where resonances are treated according
to the formalism of ref.~\cite{Jezo:2015aia} is also in preparation~\cite{JLNOP}. In the following
I will present preliminary results obtained with this last generator in comparison with the generator
of ref.~\cite{Campbell:2014kua}.
In fig.~\ref{fig:ttbprodanddec},
I compare the two generators results for the invariant mass of the $W^+$-$\; b_{\rm jet}$ system, and for
the $B$ meson fragmentation function and transverse momentum distribution (relative to the recoiling $W$ boson)
in the top rest frame.
We see that the description of the $W^+$-$\; b_{\rm jet}$ mass distribution is very similar
in the two generators. We notice instead sensible differences in the $B$ fragmentation function. The transverse momentum
plot on the right suggests that these differences may have to do with the description of gluon emission for gluon energies
near a GeV. They could thus be due to genuine interference effects between emission in production and in decay,
or to the way in which radiation is assigned to either production or decay in the formalism of ref.~\cite{Jezo:2015aia}.
This assignment becomes in fact ambiguous for gluon energies of the order of the top width. Further studies are needed
to clarify this issue.
\section{Conclusions and observation on the top mass measurement}
In this brief review I have focused on recent progress in theoretical calculations
relevant to the physics of top production at the LHC.
This progress will be vital for the top physics program at the LHC, both
for search of new physics effects and for testing our understanding
of collider physics in general.
There is however one specific topic for which this year's theoretical progress is
particularly relevant, i.e. the top mass measurement.

It has been argued that, since the top mass is extracted by fitting Monte Carlo computed distributions
to experimental data, what is really measured
is a Monte Carlo parameter, a fact that was also reminded in the experimental talk
on top mass measurements at this conference~\cite{ACastro}.
On the theory side, several arguments supporting this view are given.
It is often claimed, for example, that unless the generators being used have NLO accuracy in both top
production and decays, one cannot claim that the measurement is extracting a theoretically
well-defined top mass parameter~\cite{Corcella}. It has also been argued that,
since soft radiation from top quark is simulated
by general purpose shower Monte Carlo, that typically lack a rigorous treatment of the boundary
between perturbative and non-perturbative effects, its mass definition does not corresponds to a well
defined field theoretical one~\cite{Hoang:2014oea}.
While (on the positive side) these claims may lead to a more critical view an what is actually
measured, they are in fact unnecessary once the
theoretical error in the simulation of the top production and decay is properly
considered. As a simple example, we can consider
two kinds of determination of the top mass: one obtained from its total production
cross section, and one obtained by fitting observables related to the mass of $W$-$\;b_{\rm jet}$ system.
In the first case, assuming that we didn't have any better calculation,
we could compute the cross section using a general purpose Monte Carlo,
and fit its normalization to the data by adjusting its top mass parameter.
Once we include the associated
theoretical uncertainty (obtained for example by scale variation), we would find
a large variation of order $\as m_t$. It is certainly true that, since the generator
we are using is only accurate at leading order, we would not be able to distinguish between the \MSB{}
and pole mass, since their difference is of order $\as m_t$. However, the estimated
error itself would lead to the same conclusion. Rather than concluding that we are measuring a
``Monte Carlo'' mass, we would conclude that, because of the large error, we cannot distinguish
among the two mass definitions.

In the second example, since we
fit observables related to the mass of the $W$-$\;b_{\rm jet}$ system, we are clearly
sensitive the top pole mass, that has no NLO corrections by definition.
Of course, the relation
of the reconstructed mass to the underlying parameter in the Lagrangian
still has perturbative and non-perturbative errors that should be estimated,
but this does not mean that we are not measuring the pole mass. The error estimate
in our model of production and decay would clarify how close we are, theoretically
to the pole mass.

The issue of the interplay of soft radiation from top and non-perturbative effect
is certainly more subtle. It should not be
forgotten, however, that Shower Monte Carlo's do cut off gluon radiation from top at
scales of order of its width, and this suppression also emerges naturally
in rigorous perturbative calculation that include finite width effects.
The boundary
between perturbative and non-perturbative effects is thus more likely to be approached
by soft radiation from the top decay products, rather than from the top itself, and it
is implausible to ascribe the associated
error to the pole mass definition in the Monte Carlo.

The presence of the infrared renormalon in the top pole mass, and the
fact that (since the top is coloured) there is no unambiguous definition
of a reconstructed top, have lead many authors to pessimistic conclusions regarding the
top mass measurement, sometimes summarized by the statement that
a further error of the order of 1~GeV should be always ascribed to the so called
``Monte Carlo mass''.
As we have seen, in reality the error due to the renormalon is much smaller than
1~GeV. The inability to define a reconstructed top even at the theoretical
level is instead a serious problem, that is certainly bound to yield an irreducible
theoretical error of the order of few hundred MeV on the top mass. The real question is
whether ``few'' is one or ten.

The standard methods for measuring the top mass (reviewed in~\cite{ACastro}
at this conference) make use of a reconstructed top definition comprising
in essence the $W$-$\;b_{\rm jet}$ system.
Assuming that we have a perfect detector, and can measure exactly the mass
distribution of $W$-$\;b_{\rm jet}$ system, the question is how to relate this distribution
to the fundamental parameter of the Standard Model Lagrangian.
As we have seen in this conference,
we do have generators that include radiative corrections in top production and
decays. Furthermore, generators that also
include interference effects arising from radiation in production and decays
are around the corner. These generators can and should be used to estimate the perturbative
error on the extracted top pole mass, by considering, as usual, different scale
choices and variations of other parameters in the generators. Comparison
among different generators of the same accuracy
can also be used to better assess the error.

Estimating the error due to hadronization and non-perturbative effects
is a more difficult task,
since we do not have a rigorous theoretical framework for modeling
these effects.
 This problem, however, is not new, and not
specific to top production. In the early times of jet physics at LEP,
studies on the determination of the strong coupling constant where
performed by correcting shape variables for hadronization effects,
carried out with general purpose Shower Monte Carlo. These corrections
were leading to more consistent values in the determination of the
strong coupling constant obtained from different shape variables.
At that time, comparison of the hadronization corrections obtained with
Herwig and Pythia were used to assess the hadronization error.
A similar attitude can be adopted for the top mass measurement, although it is
certainly recommendable to work harder on the error estimate.
As a general
indication, different physically motivated hadronization schemes that
yield reasonable fits to hadronization data should be tried and used
for this purpose.
Studies like the ones presented by Corcella in this
conference~\cite{Corcella}, where the Monte Carlo performance in case
of a fictitious long-lived top is compared to the realistic case, and
like those presented in refs.~\cite{Argyropoulos:2014zoa,Christiansen:2015yqa}, where
alternative colour reconnection models are considered, should be
critically considered and extended. More ideas and research along these
lines would be welcome.

There are several proposals of alternative observables to be used for
the determination of the top mass. Some of them have been already used
by the experimental collaborations, and have been reviewed at this
conference~\cite{MarcelVOS}.  These proposals are all valuable, and it
is likely that in the future they may reach competitive precisions.
It is likely that, at the end, consistency between different
determinations will yield a convincing assessment of the overall
error.  It should not be forgotten, however, that, as the precision
increases, more and more effort should be made in trying to estimate
and reduce the theoretical error, and it is unlikely that this task
will be easier for these alternative methods than with the current
main-stream approach.

\end{document}